\documentclass[aps,pre,twocolumn,superscriptaddress,showkeys]{revtex4-2}

\usepackage{graphicx}
\usepackage{amsmath}
\usepackage{amssymb}
\usepackage{xcolor}
\usepackage{tikz}
\usepackage{siunitx}

\definecolor{darkred}{rgb}{0.6350,0.0780,0.1840}
\definecolor{softdarkred}{rgb}{0.75,0,0}
\definecolor{darkgreen}{rgb}{0,0.5,0}
\definecolor{darkyellow}{rgb}{0.9290,0.6940,0.1250}
\definecolor{darkorange}{rgb}{0.8500,0.3250,0.0980}
\definecolor{darkblue}{rgb}{0,0.4470,0.7410}
\definecolor{darkred}{rgb}{0.6350,0.0780,0.1840}
\DeclareRobustCommand\fullblack {\tikz[baseline=-0.6ex ]\draw[thick] (0,0)--(0.4,0);}

\DeclareRobustCommand\fullred {\tikz[baseline=-0.6ex ]\draw[draw=red,thick] (0,0)--(0.4,0);}
\DeclareRobustCommand\fullred{\tikz[baseline=-0.6ex ]\draw[red,thick] (0,0)--(0.4,0);}
\DeclareRobustCommand\dashedblack {\tikz[baseline=-0.6ex ]\draw[thick,dashed] (0,0)--(0.4,0);}

\DeclareRobustCommand\dottedblack {\tikz[baseline=-0.6ex ]\draw[thick,dotted] (0,0)--(0.4,0);}
\DeclareRobustCommand\dashedred {\tikz[baseline=-0.6ex ]\draw[red,thick,dash dot] (0,0)--(0.4,0);}

\begin{document}


\title{Collapse dynamics of dry granular columns: from free-fall to quasi-static flow}


\author{Wladimir Sarlin}
\email[]{wladimir.sarlin1@universite-paris-saclay.fr}
\affiliation{Universit\'e Paris-Saclay, CNRS, Laboratoire FAST, F-91405 Orsay, France}
\author{Cyprien Morize}
\email[]{cyprien.morize@universite-paris-saclay.fr}
\affiliation{Universit\'e Paris-Saclay, CNRS, Laboratoire FAST, F-91405 Orsay, France}
\author{Alban Sauret}
\affiliation{University of California, Santa Barbara, Department of Mechanical Engineering, CA 93106, USA}
\author{Philippe Gondret}
\affiliation{Universit\'e Paris-Saclay, CNRS, Laboratoire FAST, F-91405 Orsay, France}


\date{\today}

\begin{abstract}
Gravity-driven collapses involving large amounts of dense granular material, such as landslides, avalanches, or rockfalls, in a geophysical context, represent significant natural hazards. Understanding their complex dynamics is hence a key concern for risk assessment. In the present work, we report experiments on the collapse of quasi-two-dimensional dry granular columns under the effect of gravity, where both the velocity at which the grains are released and the aspect ratio of the column are varied to investigate the dynamics of the falling grains. At high release velocity, classical power laws for the final deposit are recovered, meaning those are representative of a free-fall like regime. For high enough aspect ratios, the top of the column undergoes an overall free-fall like motion. In addition, for all experiments, the falling grains also spread horizontally in a free-fall like motion, and the characteristic time of spreading is related to the horizontal extension reached by the deposit at all altitudes. At low release velocity, a quasi-static state is observed, with scaling laws for the final geometry identical to those of the viscous regime of granular-fluid flow. The velocity at which the grains are released governs the collapse dynamics.
Between these two asymptotic regimes, the higher the release velocity, the smaller the impact on the collapse dynamics. The criterion $\overline{V} \geq 0.4\sqrt{gH_0}$, where $H_0$ is the initial height of the column, is found for the mean release velocity $\overline{V}$ not to influence the granular collapse.
\end{abstract}


\maketitle


\section{Introduction}

Geophysical gravity-driven flows such as landslides or avalanches are as fascinating as complex due to their inherent unsteadiness and the large deformation experienced by the flowing mass during its motion. Understanding these phenomena is of great interest as they pose serious threats to human activity in mountainous areas \cite{2015_langlois}.\\
A common approach to characterize these flows, at the laboratory scale, consists of using granular materials to mimic the slumping mass. In particular, the collapse of a column of grains is a simple but relevant configuration for modeling landslides \cite{2006_lajeunesse}, so that it has been extensively studied in the last two decades \cite{2004_lajeunesse,2005_lajeunesse,2004_lube,2005_lube,2005_staron,2005_balmforth,2005_zenit,2006_meriaux,2006_larrieu,2007_staron,2007_thompson,2008_lacaze,2010_trepanier,2011_lagree,2011_rondon,2012_tapia-mcclung,2012_topin,2013_artoni,2013_degaetano,2014_warnett,2015_mutabaruka,2015_ionescu,2018_jing,2018_bougouin,2019_cabrera,2021_ordaz,2021_man,2021_sun,2021_yang}. The aim of these studies was to reach a better understanding of such flows and describe the final morphology of the deposits. The usual experimental configuration consists of a column of granular material of height $H_0$ and initial width $L_0$ (two-dimensional setup \cite{2005_lajeunesse}) or radius $R_0$ (axisymmetric setup \cite{2004_lube}), initially at rest. When the column is released, the grains fall and spread over the ground. The initial aspect ratio of the column, defined as $a=H_0/L_0$, was found to govern the final geometry of the deposits. The influence of $a$ on the final height $H_f$ and runout length $\Delta L_f=L_f-L_0$ has been captured through power laws both for the rectangular \cite{2005_lube,2005_lajeunesse} and the axisymmetric \cite{2004_lube,2004_lajeunesse} geometries.
For instance, Lajeunesse \textit{et al.} \cite{2005_lajeunesse} found that for glass beads in a rectangular channel the relative runout distance $\Delta L_f/L_0$ was proportional to $a$ when $a \lesssim 3$ and to $a^{2/3}$ when $a \gtrsim 3$, while the relative final height $H_f/L_0$ was equal to $a$ for $a \lesssim 0.7$ and proportional to $a^{1/3}$ at higher aspect ratios. In the axisymmetric configuration, different scaling laws have been obtained \cite{2004_lube,2004_lajeunesse,2005_lajeunesse}.\\ 
Numerical simulations reproduced these scalings for the final geometry using Contact Dynamics algorithms \cite{2005_staron}, shallow water equations \cite{2006_larrieu}, or continuum approaches \cite{2011_lagree,2015_ionescu} implementing pressure-dependent granular rheologies such as the $\mu(I)$ rheology introduced by Jop \textit{et al.} \cite{2006_jop}.
The numerical values of the prefactors, exponents, and critical aspect ratios of these power laws slightly vary between authors \cite{2005_lajeunesse,2005_lube,2005_balmforth,2005_staron,2018_jing,2019_cabrera}.
In particular, the exponents were shown to be independent of the material properties, which only affect the numerical prefactors \cite{2007_staron,2005_balmforth,2021_man}. However, despite extensive works, no clear explanation exists to rationalize these scalings.\\
Experiments conducted by Mériaux \cite{2006_meriaux} addressed the case where inertia is negligible in the problem, \textit{i.e.}, when the granular column is slowly released using a horizontally moving gate. In this situation, different empirical scalings are obtained, which differ from those obtained when the column is instantaneously released \cite{2005_staron}. In a different context, the collapse of a liquid-immersed granular column was investigated both experimentally \cite{2011_rondon,2018_bougouin,2021_sun,2021_yang} and numerically \cite{2012_topin,2018_jing,2021_yang}. In this case, the collapse dynamics depends not only on the aspect ratio $a$ of the column, but also on the density ratio between the granular medium and the surrounding fluid, and the Stokes number, which compares the grain inertia to the viscous fluid forces. In particular, the viscous regime, at low Stokes number, is characterized by the absence of grain inertia \citep{2018_bougouin}.\\
Nevertheless, whereas extensive efforts were made to investigate the behavior of the runout distance at the base of the column, there is a lack of experimental investigations focusing on the overall dynamics of the collapse. In the present work, the granular slumping dynamics is investigated in detail by varying the velocity at which the grains are released for different aspect ratios of the initial granular column. The experimental setup is first described in section \ref{SecII}. In section \ref{SecIII}, the different regimes observed depending on the release velocity are characterized qualitatively. Quantitative results are discussed in section IV, where we describe both the dynamics at high and low release velocity as well as the transition between these two asymptotic regimes.

\section{Experimental setup}
\label{SecII}

\begin{figure}[t]
	\centering
	\includegraphics[width=\columnwidth]{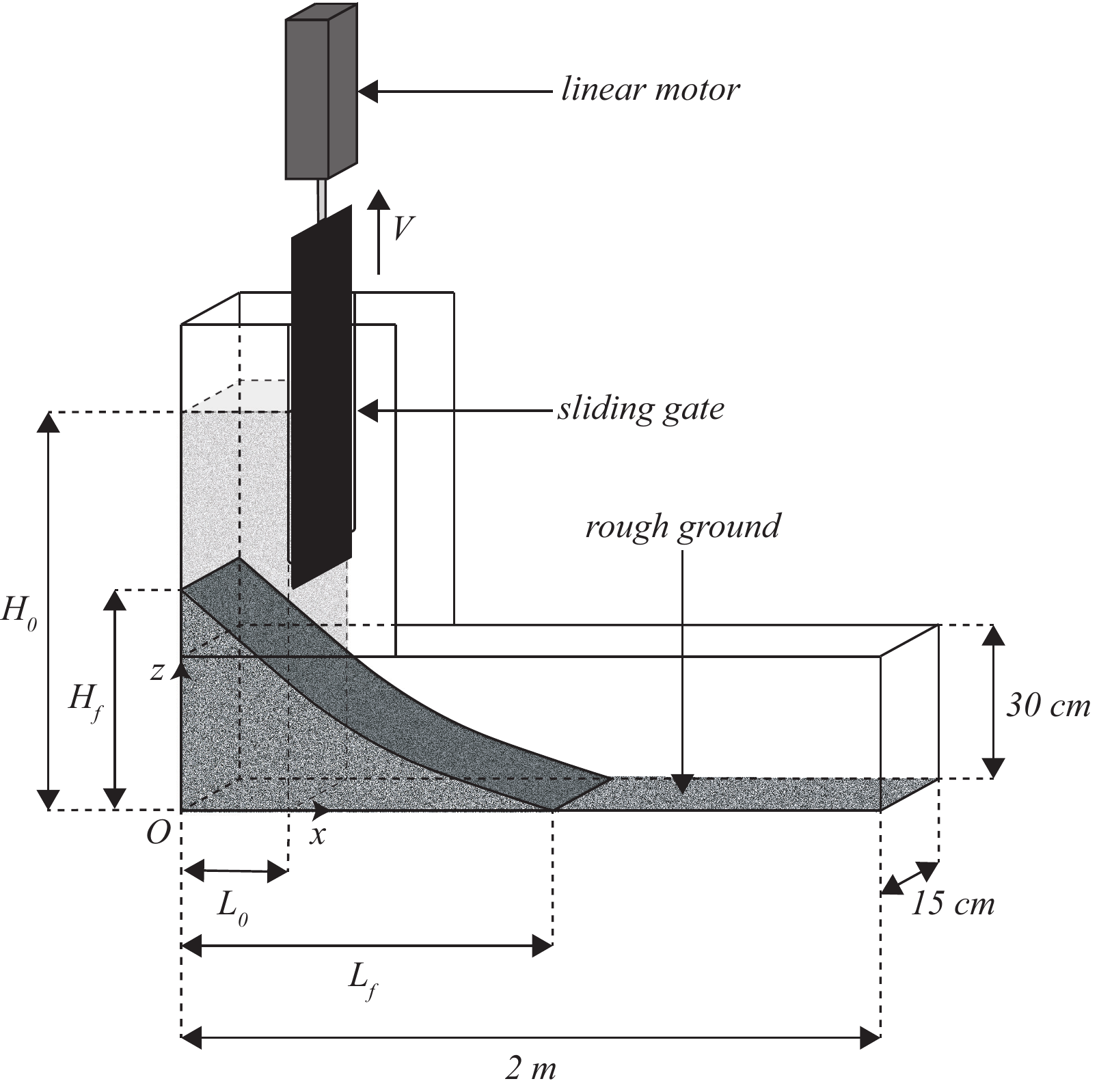}
	\caption{Sketch of the experimental setup, showing the initial rectangular granular column of height $H_0$ and width $L_0$, and a schematic view of the final deposit of height $H_f$ and runout length $L_f$.}
	\label{fig:fig1}
\end{figure}

The experiments were conducted using the setup presented in figure \ref{fig:fig1}. On the left side of a  $2\ \rm{m}\times0.15\ \rm{m}\times0.3\ \rm{m}$ parallelepipedic glass tank, a column of granular material is initially retained by a sliding vertical gate, located at a distance $L_0$ from the left wall. The $x$-axis is along the horizontal direction, while the $z$-axis is along the vertical one. The origin is located at the bottom left end of the experimental setup. A flat rough ground, made with the same grains as the granular column, covers the bottom of the tank to ensure a no-slip boundary condition. To avoid possible segregation effects due to polydispersity \cite{2013_degaetano}, we used monodisperse glass beads of diameter $d \simeq 5$~mm and density $\rho \simeq $ 2.5 g/cm$^3$, with a measured packing fraction $\phi \simeq 0.64$. Besides, the angle of repose $\theta_r$ was measured at about $23.5 \pm 1.2^\circ$, in agreement with previous studies \cite{2005_lajeunesse,2014_sauret}. The spanwise dimension of the channel was chosen large enough to prevent confinement effects \cite{2002_courrech_du_pont}. It was indeed verified that a channel width greater than 10 cm, corresponding to 20$d$, ensures no significant influence of the sidewalls on the granular collapse \cite{2019_robbe-saule} [see figure 2.18 herein].\\
At $t=0$, the gate is lifted using a brushless servomotor so that the column collapses and spreads under the effect of gravity. Using a motor allows controlling the nominal velocity $V$ at which the column is released. Across experiments, this release velocity was varied over three decades, namely from $1\ \rm{mm.s^{-1}}$ to  $1.2\ \rm{m.s^{-1}}$, for different initial aspect ratios of the column.

\setlength{\tabcolsep}{10pt}
\begin{table}[t]
  \begin{center}
  \begin{tabular}{ccccc}
  \hline
   & $H_{0}$ (cm) & $L_{0}$ (cm) & $a$ & $V$ (m/s) \\
 \hline
 A & [10 - 50] & [2.5 - 20] & [0.5 - 20] & 1.2\\ 
 B & [6 - 45] & [7.5 - 20] & [0.3 - 6] & \num{1.0e-2}\\
 C & 15 & 20 & 0.75 & [\num{1.0e-3} - 1.2]\\ 
 D & 20 & 10 & 2 & [\num{1.0e-2} - 1.2]\\
 E & 37.5 & 7.5 & 5 & [\num{1.0e-3} - 1.2]\\  
 \hline
  \end{tabular}
  \caption{Experimental parameters considered in this study: height $H_0$, width $L_0$, and aspect ratio $a$ of the initial granular column, and nominal release velocity $V$ of the sliding gate.}
  \label{tab:table1}
  \end{center}
\end{table}

\noindent The parameters considered in this study are reported in Table \ref{tab:table1}, and are divided into five series. The experiments conducted at high release velocity ($V = 1.2\ \rm{m.s^{-1}}$) are gathered in Series A, while experiments  performed at a low release velocity ($V = 1\ \rm{cm.s^{-1}}$) are collected in Series B. Finally, the transition between these two limits was investigated in series C, D and E, each corresponding to a fixed aspect ratio $a$ of 0.75, 2, and 5, respectively. The corresponding values for $H_0$ and $L_0$ are also reported in Table~\ref{tab:table1}.\\
The collapse dynamics is recorded from the sidewall of the tank using a Nikon D3300 camera, operating at $50\ \rm{Hz}$. Image sequences are then processed to obtain the time evolution of the granular contour using a custom-made MATLAB routine based on a thresholding method. 

\section{Phenomenology}
\label{SecIII}

Different collapse dynamics and geometries of the final deposit are observed when varying the release velocity. Two asymptotic behaviors can be highlighted for high and low values of $V$. In the first situation, corresponding to the experiments conducted at high release velocity (series A), the classical granular collapse is observed, as previously reported in the literature \cite{2005_lube,2005_lajeunesse,2005_staron}. The grains quickly fall and spread over the ground with high inertia, as illustrated in figure \ref{fig:fig2}(a)-(f) for a column with a high aspect ratio ($a=5$) and, to a lesser extent, in figure \ref{fig:fig3}(a)-(e) in the case of a low aspect ratio ($a=0.5$). While the bottom of the column mainly follows a horizontal motion [figures \ref{fig:fig2}(c) and \ref{fig:fig3}(c)], its top seems to undergo a vertical acceleration for high enough aspect ratios ($a \gtrsim 3$) [top of the column in figures \ref{fig:fig2}(b) and \ref{fig:fig2}(c)]. This behavior was identified as the free-fall regime by Staron and Hinch \cite{2005_staron}. The overall motion presents strong unsteadiness and leads to a final geometry that exhibits a significant curvature of the deposit, especially at large aspect ratio [figure \ref{fig:fig2}(f)].\\
The experiments conducted at low release velocity (Series B) show a very different behavior, as can be seen in figures \ref{fig:fig2}(g)-(l) and \ref{fig:fig3}(f)-(j). A quasi-static flow occurs, and at all times the moving granular contour is roughly triangular, with a foot angle close to the angle of repose $\theta_r$ of the material. Only minor deviations from a triangular shape are observed, especially at the onset of the granular slide, as illustrated by the slightly curved interface in figures \ref{fig:fig2}(h)-(j).
In that respect, the final morphology of the deposit has no significant curvature at the end of the slide, and for large enough aspect ratios ($a \gtrsim 0.8$) it exhibits a triangular shape, as illustrated in figure \ref{fig:fig2}(l). At a given aspect ratio, the runout distance is systematically larger for experiments at large release velocity, while the final height is higher for experiments at low release velocity. For low enough aspect ratios ($a \lesssim 0.8$), a trapezoidal shape is observed for the final deposit as can be seen in figure \ref{fig:fig3}(j), since only a fraction of the initial column collapses. In that case, the runout distance is again larger for the experiments at large release velocity, while the final height systematically coincides with the initial one.\\
Between these two asymptotic situations of high and low release velocity, a transition regime is observed: for a given value of $a$, when the release velocity is increased, the runout distance and the curvature of the final deposit surface increase, while the final height decreases (when $a\gtrsim0.8$) or stays constant (for $a\lesssim0.8$).

\section{Results and Discussion}
\label{SecIV}

\subsection{Free-fall regime}
\label{SubsecIVA}

The final height $H_f$ and runout distance $\Delta L_f = L_f-L_0$ of the deposit (with $H_f$ and $L_f$ defined as in figure \ref{fig:fig1}) are systematically determined for each experiment. It should be pointed out that $L_f$ is actually evaluated at $z=d$, \textit{i.e.}, at one grain diameter from the bottom plate, to reduce measurement uncertainties.
The evolution of the relative final height $H_f/L_0$ and runout distance $\Delta L_f/L_0$ with the initial aspect ratio is presented in figures \ref{fig:fig4}(a) and \ref{fig:fig4}(b), respectively, for the experiments at large release velocity ($\bigstar$). In addition, the following fits, inspired by Lajeunesse \textit{et al.} \cite{2005_lajeunesse}, for the relative runout distance,

\begin{equation}
	\frac{\Delta L_f}{L_0} \simeq \left\{ \begin{array}{ll}
	\displaystyle 1.85~a\\[8pt]
	\displaystyle 2.67~a^{2/3}
	\end{array}\right.
    \begin{array}{ll}
	\displaystyle \ \mathrm{for\ } a \lesssim 3,\\[8pt]
	\displaystyle \ \mathrm{for\ } a \gtrsim 3,
	\end{array}
  \label{lajeunesse_runout}
\end{equation}

\noindent and final height,

\begin{equation}
	\frac{H_f}{L_0} \simeq \left\{ \begin{array}{ll}
	\displaystyle a\\[8pt]
	\displaystyle 0.93~a^{1/3}
	\end{array}\right.
    \begin{array}{ll}
	\displaystyle \ \mathrm{for\ } a \lesssim 0.8,\\[8pt]
	\displaystyle \ \mathrm{for\ } a \gtrsim 0.8.
	\end{array}
  \label{lajeunesse_final_height}
\end{equation}

\begin{figure}[t]
	\centering
	\includegraphics[width=\linewidth]{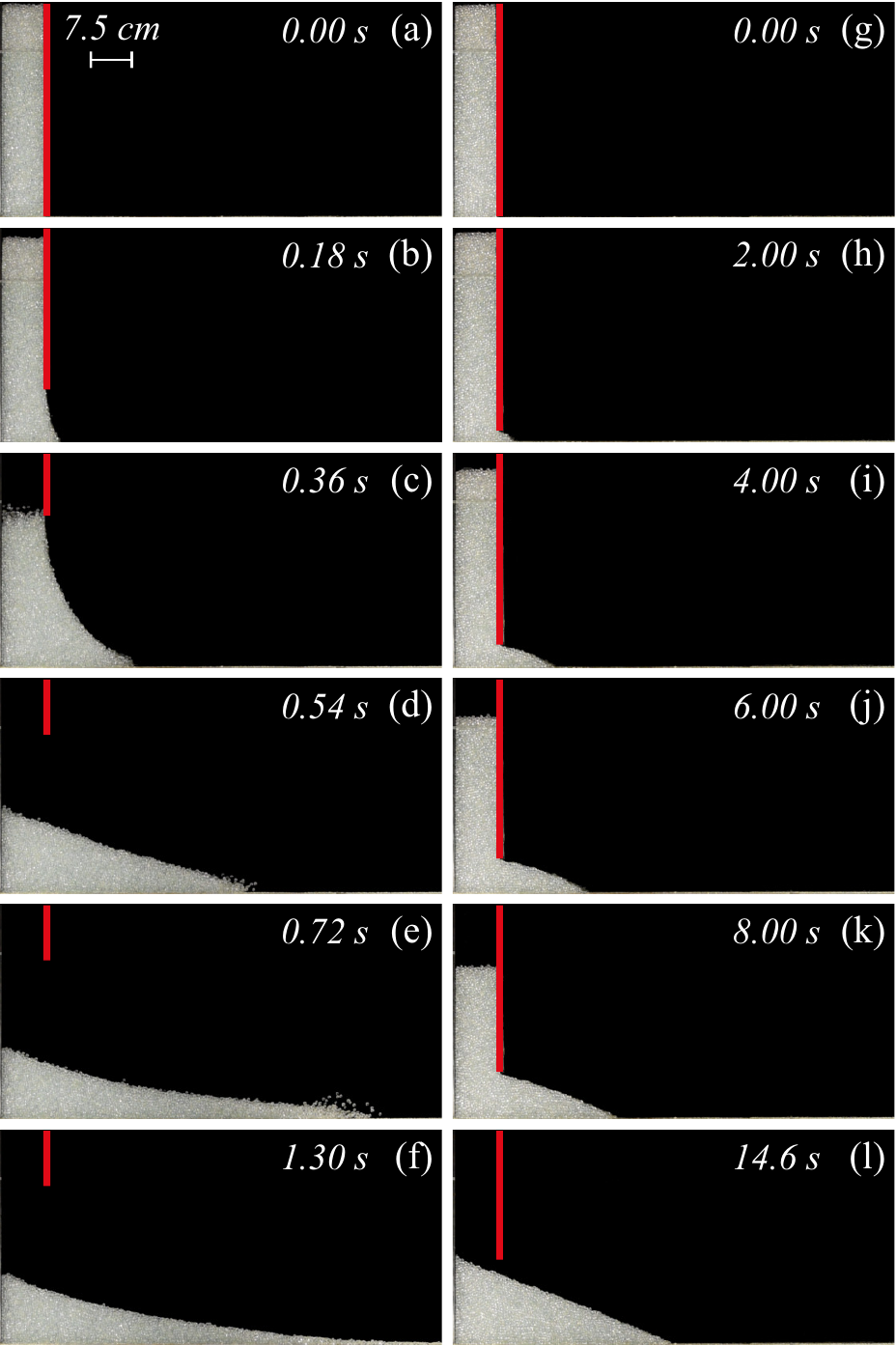}
	\caption{Image sequences of the collapse of a rectangular granular column with $H_0=37.5\ \rm{cm}$ and $L_0=7.5\ \rm{cm}$ ($a=5$) (a)-(f) for high release velocity ($V=1.2\ \rm{m.s^{-1}}$), and (g)-(l) for low release velocity ($V=0.01\ \rm{m.s^{-1}}$). The red thick line indicates the location of the sliding gate.}
	\label{fig:fig2}
\end{figure}

\begin{figure}[t]
	\centering
	\includegraphics[width=\linewidth]{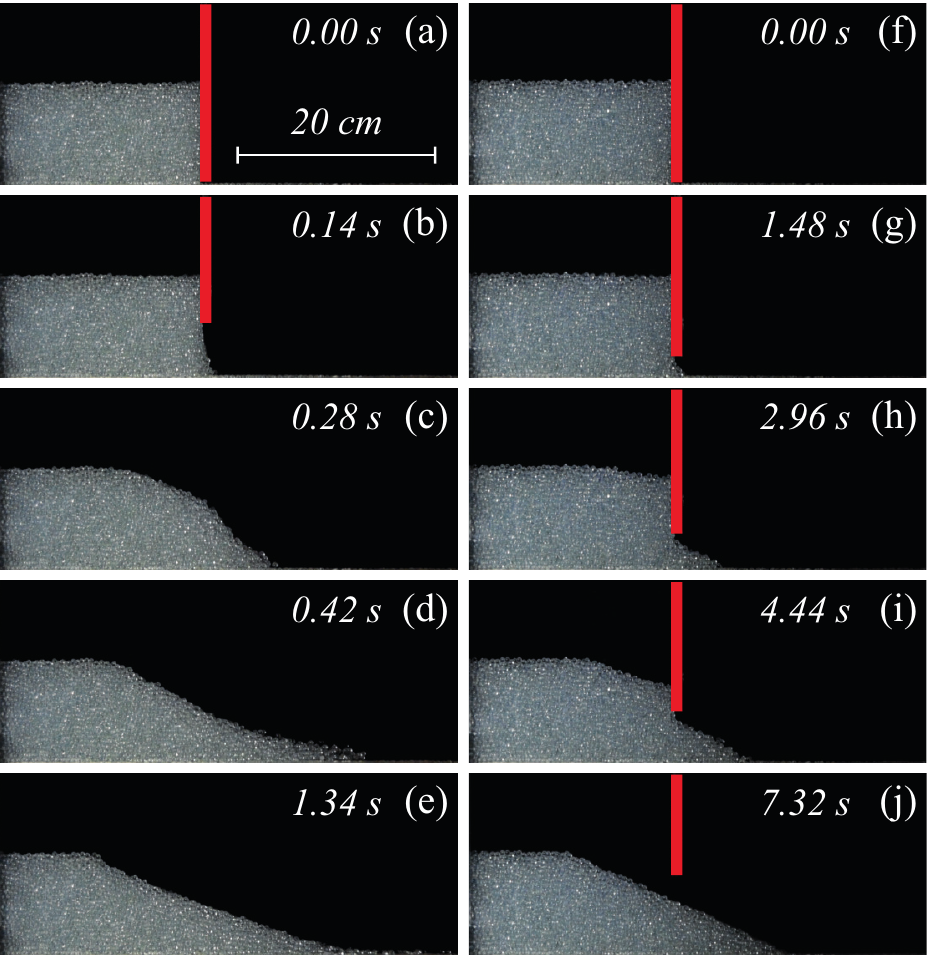}
	\caption{Image sequences of the collapse of a rectangular granular column with $H_0=10\ \rm{cm}$ and $L_0=20\ \rm{cm}$ ($a=0.5$) (a)-(e) for high release velocity ($V=1.2\ \rm{m.s^{-1}}$), and (f)-(j) for low release velocity ($V=0.01\ \rm{m.s^{-1}}$). The red thick line indicates the location of the sliding gate.}
	\label{fig:fig3}
\end{figure}

\noindent are also reported (black dashed lines). An excellent agreement is observed between the experimental data obtained at high release velocities and the scaling laws given by \eqref{lajeunesse_runout} and \eqref{lajeunesse_final_height}.\\
To investigate in more detail the dynamics of the collapse, we observe the time evolution of the height $H(x,t)$ for different values of $x$, between $0$ and $L_0$. In addition, the time evolution of the spreading length $L(z,t)$ is also extracted at different altitudes $z$, between $0$ and $H_f$, for all aspect ratios considered in this study. An example of the time evolution of these parameters is reported in figures \ref{fig:fig5}(a)-(b), for an initial column with $H_0=37.5\ \rm{cm}$ and $L_0=7.5\ \rm{cm}$ ($a=5$). In figure \ref{fig:fig5}(a), $H(x,t)$ is plotted as a function of $t-t_{0z}$, where $t_{0z}$ corresponds to the time for which a vertical displacement greater than $1.5d$ is detected. $H(x,t)$ decreases from $H_0$ to a final height $H_f(x)$, as presented in figure \ref{fig:fig5}(a) for ten equally-spaced values of $x$ between $0$ and $L_0$. An acceleration phase, followed by a deceleration stage, is observed. During the acceleration period, all experimental data collapse on a master curve, whose equation corresponds to a free-fall like motion,

\begin{equation}
	H(x,t) \simeq H_0 - \frac{1}{2} \alpha g \left( t-t_{0z} \right)^2,
  \label{collapse_law_of_motion}
\end{equation}

\begin{figure}[t]
	\centering	\includegraphics[width=\columnwidth]{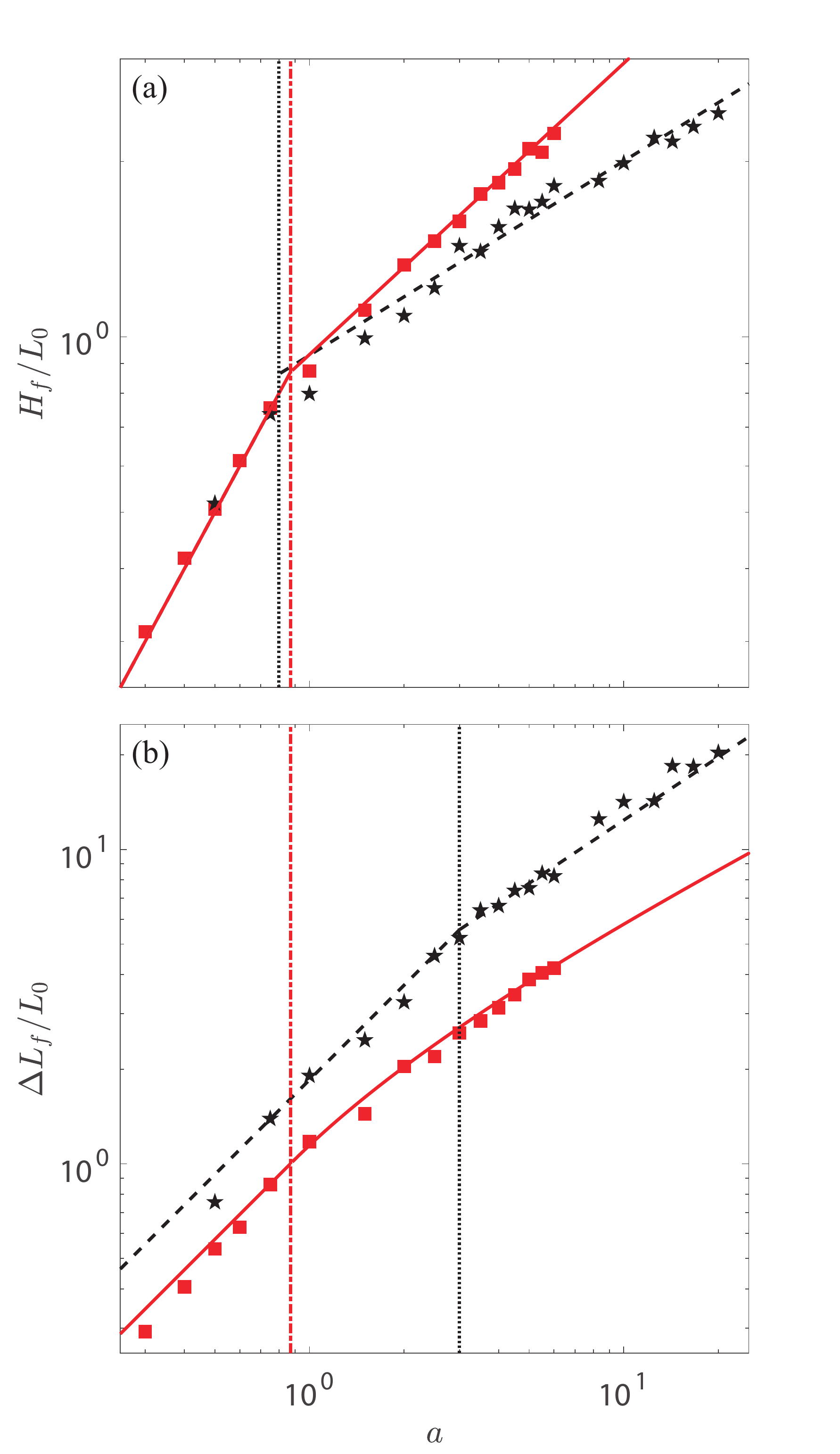}
	\caption{Evolution of (a) the relative final height $H_f/L_0$ and (b) the relative runout distance $\Delta L_f/L_0$ of the deposits as a function of the initial aspect ratio $a$ of the column. ($\bigstar$) Data from series A; (\textcolor{red}{$\blacksquare$}) data from series B; (\dashedblack) power laws from equations \eqref{lajeunesse_runout} and \eqref{lajeunesse_final_height}; (\fullred) equations (\ref{eq:lrs_ha_scalingsa}), (\ref{eq:lrs_ha_scalingsb}), (\ref{eq:lrs_la_scalingsa}), and (\ref{eq:lrs_la_scalingsb}), with $\theta_r = 23.5^\circ$; (\dottedblack) critical aspect ratios of the free-fall regime (a) $a \simeq 0.8$ and (b) $a \simeq 3$ from equations \eqref{lajeunesse_runout} and \eqref{lajeunesse_final_height}, respectively; (\dashedred) critical aspect ratio of the quasi-static regime $a_{cr} \simeq 0.87$, as defined in equation \eqref{eq:acr}.}
	\label{fig:fig4}
\end{figure}

\noindent where $\alpha \simeq 0.56$. As illustrated in figure \ref{fig:fig5}(a), the different curves depart from this law of motion when the deceleration stage of the collapse occurs, here around $0.25$ s. This separation happens at a slightly different time for each curve, and leads to different final heights depending on the considered value of $x$.
\\In figure \ref{fig:fig5}(b), the spreading length $L(z,t)$ is plotted, for ten equally-spaced altitudes between $0$ and $H_f$, as a function of $t-t_{0x}$, where $t_{0x}$  is the time the bottom of the sliding gate reaches $z$, which allows initiating the spreading at this altitude. At each value of $z$, $L(z,t)$ also exhibits an acceleration phase, followed by a deceleration. At the end of the deceleration stage, $L(z,t)$ reaches an asymptotic value $L_f(z)$, after passing through a maximum value.\\ 
The acceleration phases at all altitudes also seem to collapse on a master curve, which is found to be proportional to a free-fall like motion,

\begin{figure}[t]
	\centering
	\includegraphics[width=\columnwidth]{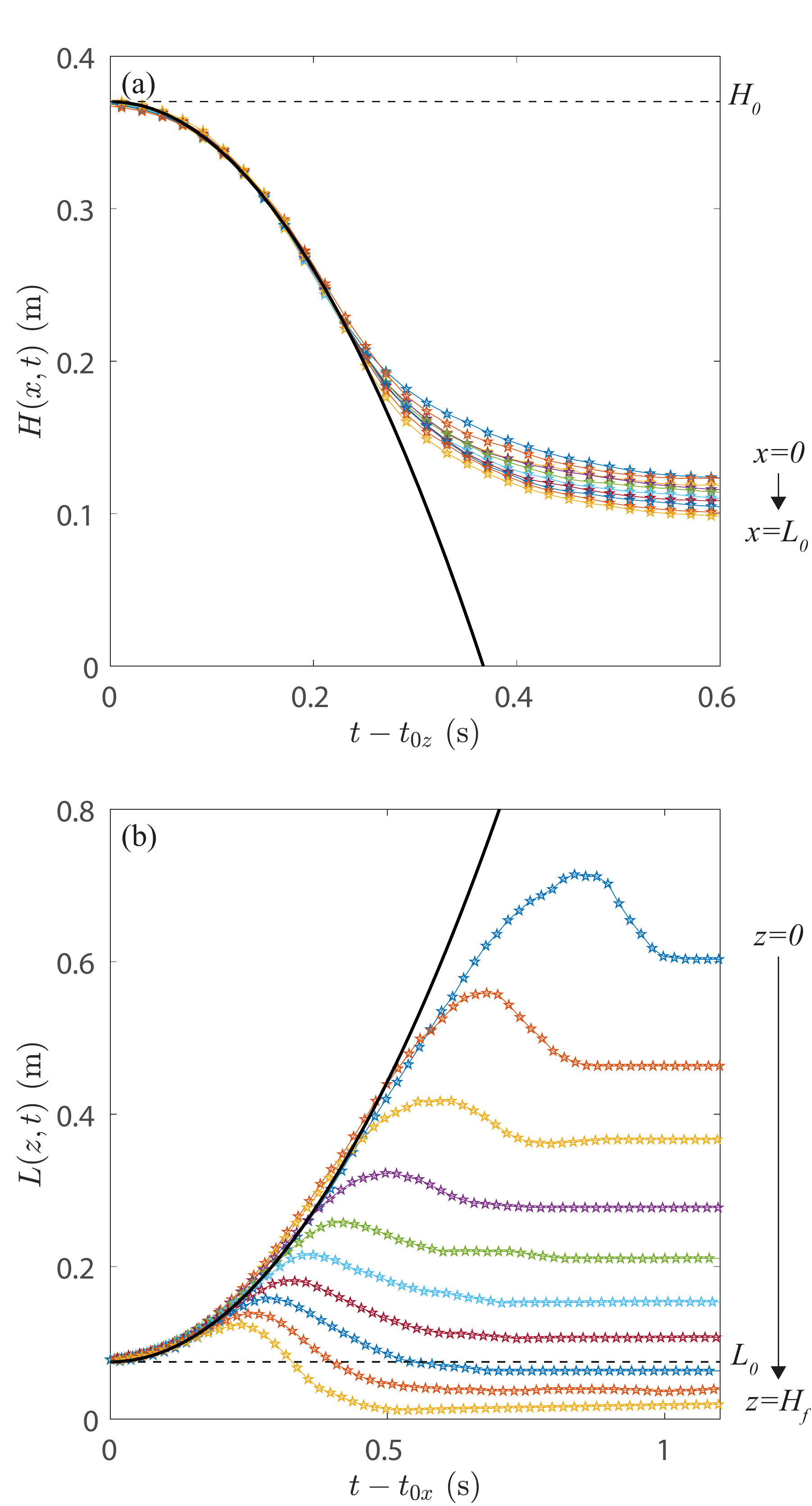}
	\caption{Example of the time evolution of (a) the height $H(x,t)$ and (b) the spreading length $L(z,t)$ of the collapse, for an initial column of height $H_0=37.5\ \rm{cm}$ and width $L_0=7.5\ \rm{cm}$ ($a=5$). The different curves correspond to ten equally-spaced values of (a) $x$ between $0$ and $L_0$ (from top to bottom) and (b) $z$ between $0$ and $H_f$ (from top to bottom); The thick lines show equation \eqref{collapse_law_of_motion} in (a), and equation \eqref{spreading_law_of_motion} in (b).}
	\label{fig:fig5}
\end{figure}

\begin{equation}
	L(z,t) \simeq L_0 + \frac{1}{2} \beta g\left( t-t_{0x} \right)^2,
  \label{spreading_law_of_motion}
\end{equation}

\noindent with $\beta \simeq 0.30$. The experimental curves deviate from equation \eqref{spreading_law_of_motion} earlier at larger altitude, revealing that the acceleration stage of the spreading is getting shorter for increasing values of $z$.

\begin{figure}[t]
	\centering
	\includegraphics[width=\columnwidth]{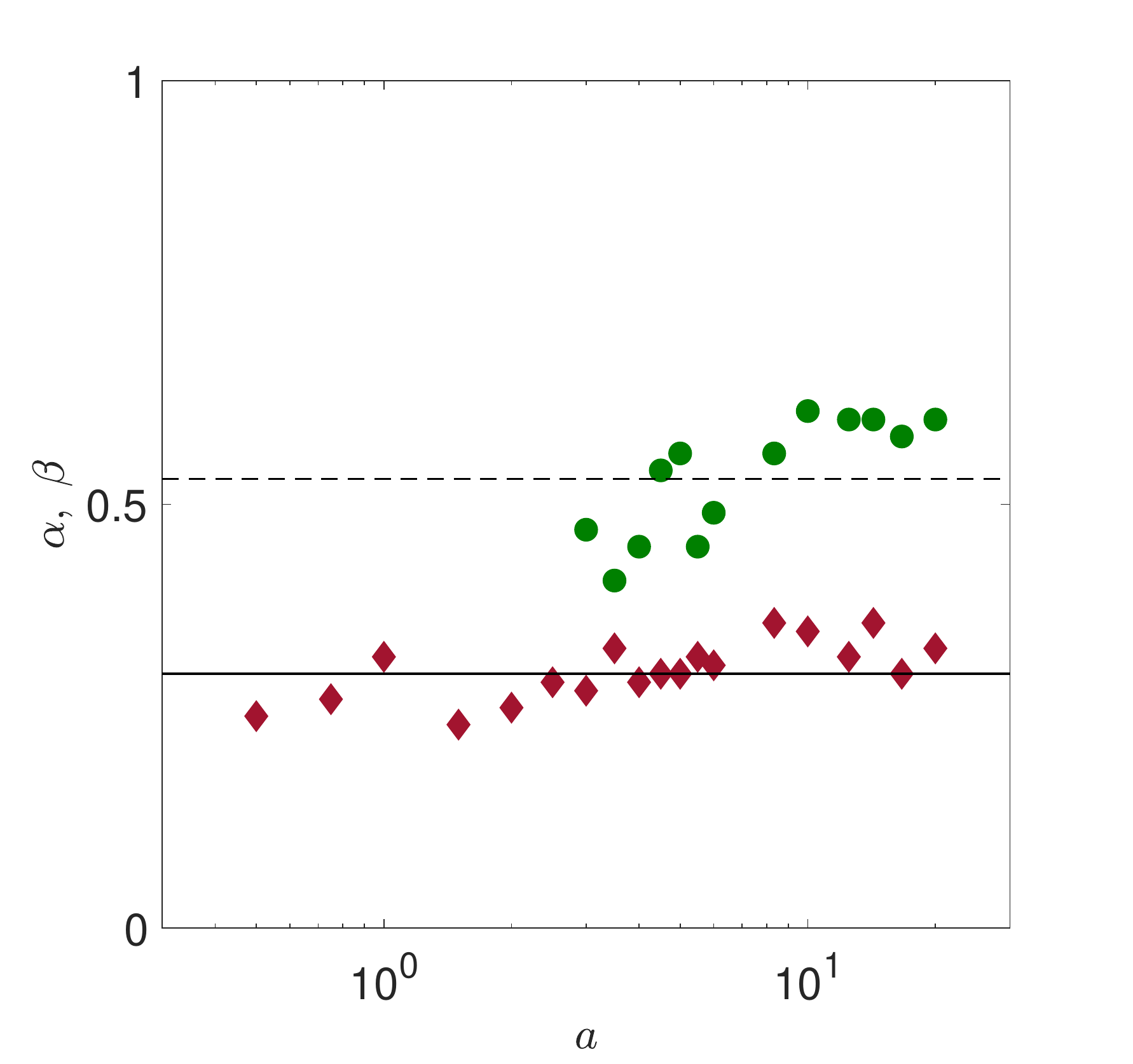}
	\caption{Evolution of the coefficients (\textcolor{darkgreen}{$\bullet$}) $\alpha$ and (\textcolor{darkred}{$\blacklozenge$}) $\beta$ with the aspect ratio $a$, at high release velocity. Mean values (\dashedblack) $\overline{\alpha}=0.53$ and (\fullblack) $\overline{\beta}=0.30$ are also indicated.}
	\label{fig:fig6}
\end{figure}

\noindent The evolution of the coefficients $\alpha$ and $\beta$ with the aspect ratio of the column $a$ is reported in figure \ref{fig:fig6} for all experiments at large release velocity, \textit{i.e.}, of series A. It is important to note that there is no available values for $\alpha$ as soon as $a \lesssim 3$, as it was not possible to distinguish experimentally a clear acceleration stage, such as the one presented in figure \ref{fig:fig5}(a), for these low aspect ratios. This observation is in agreement with results from previous numerical simulations \cite{2005_staron,2018_jing}, which showed that the free-fall of the top of the column only occurred when $a \gtrsim 2.5$. Despite some scatter of the data, the $\beta$ values are significantly lower than the $\alpha$ values. There is no significant influence of the aspect ratio on $\beta$, as all experiments are close to a mean value $\overline{\beta} \simeq 0.30 \pm 0.06$. For the coefficient $\alpha$, it can be noticed that values when $a \lesssim 8$ are slightly below those obtained at higher aspect ratios, which saturates at a value close to 0.6. However, at first order, all values are roughly distributed around a mean value $\overline{\alpha} \simeq 0.53 \pm 0.12$. This value, smaller than one, slightly differs from past numerical results of Staron and Hinch \cite{2005_staron} and Jing \textit{et al.} \cite{2018_jing}, but also from the experimental work of Balmforth and Kerswell \cite{2005_balmforth}. It could be an indication that the material properties as well as the boundary conditions at the sidewalls may have a significant influence on $\alpha$ and $\beta$.\\
\noindent These observations can be summarized briefly as follows: the top of the column uniformly undergoes a free-fall like motion at an acceleration of about $0.5g$ as soon as $a \gtrsim 3$, while the column spreads laterally, again in a ``free-fall" like motion at a typical acceleration of about $0.3g$, for all aspect ratios considered in this study.\\
\noindent To further confirm this free-fall like dynamics, two characteristic times are systematically extracted from the $H(x,t)$ and $L(z,t)$ curves: the characteristic times of vertical fall $\tau_z(x)$ and of horizontal spreading $\tau_x(z)$ are taken as the times at which $H(x,t)$ and $L(z,t)$ deviate by more than $10\%$ (\textit{i.e.}, out of the error range) from equations \eqref{collapse_law_of_motion} and \eqref{spreading_law_of_motion}, respectively. 

\begin{figure}[t!]
	\centering
\includegraphics[width=\columnwidth]{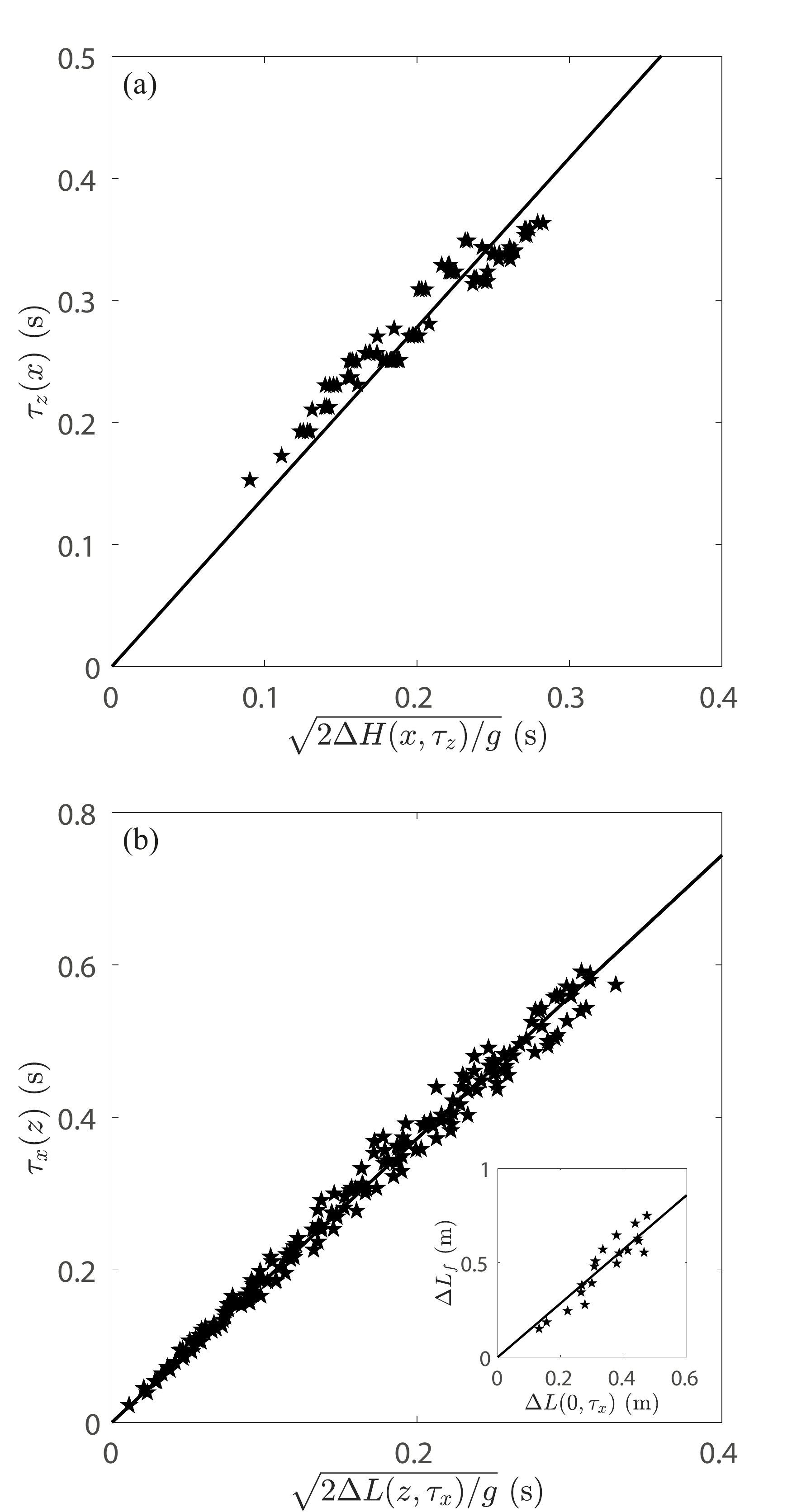}
	\caption{(a) Evolution of $\tau_z(x)$ with $\sqrt{2\Delta H(x,\tau_z)/g}$. ($\bigstar$) All experiments where $a \gtrsim 3$, with different initial aspect ratios $a$ and at different positions $x$ from the left wall. The solid line (\fullblack) is the best linear fit of slope 1.39. (b) Evolution of $\tau_x(z)$ with $\sqrt{2\Delta L(z,\tau_x)/g}$. ($\bigstar$) All experiments, for different aspect ratios $a$ and at different heights $z$ from the bottom. The solid line (\fullblack) is the best linear fit of slope 1.86. The inset shows the comparison between the final runout distance $\Delta L_f$ and the typical spreading length $\Delta L(0,\tau_x)$ at the base of the column at time $\tau_x(0)$, with (\fullblack) the best linear fit of slope 1.43.}
	\label{fig:fig7}
\end{figure}

\noindent For all experiments at large release velocity (series A), figures \ref{fig:fig7}(a)-(b) show these characteristic times $\tau_z(x)$ and $\tau_x(z)$ as functions of the typical free-fall times $\sqrt{2\Delta H(x,\tau_z)/g}$, where $\Delta H(x,\tau_z)=H_0-H(x,\tau_z)$, and $\sqrt{2\Delta L(z,\tau_x)/g}$, where $\Delta L(z,\tau_x)=L(z,\tau_x)-L_0$, respectively, at all values of $x$ (resp. $z$) considered. The best fit of the data from figure \ref{fig:fig7}(a) leads to

\begin{equation}
	\tau_z(x) \simeq 1.39\sqrt{2\Delta H(x,\tau_z)/g},
  \label{eq_tau_z_x}
\end{equation}

\noindent where the prefactor is very close to $1/\sqrt{\overline{\alpha}} \simeq 1.37$. This confirms that for tall columns the time of collapse $\tau_z(x)$ is proportional to the free-fall time over the typical variation in height $\Delta H(x,\tau_z)$. We should emphasize again that this free-fall like motion of the top of the column is absent for low aspect ratios ($a \lesssim 3$). 
In addition, figure \ref{fig:fig7}(b) shows that the relation

\begin{equation}
	\tau_x(z) \simeq 1.86\sqrt{2\Delta L(z,\tau_x)/g},
  \label{eq_tau_x_z}
\end{equation}

\begin{figure}[t]
	\centering
	\includegraphics[width=\columnwidth]{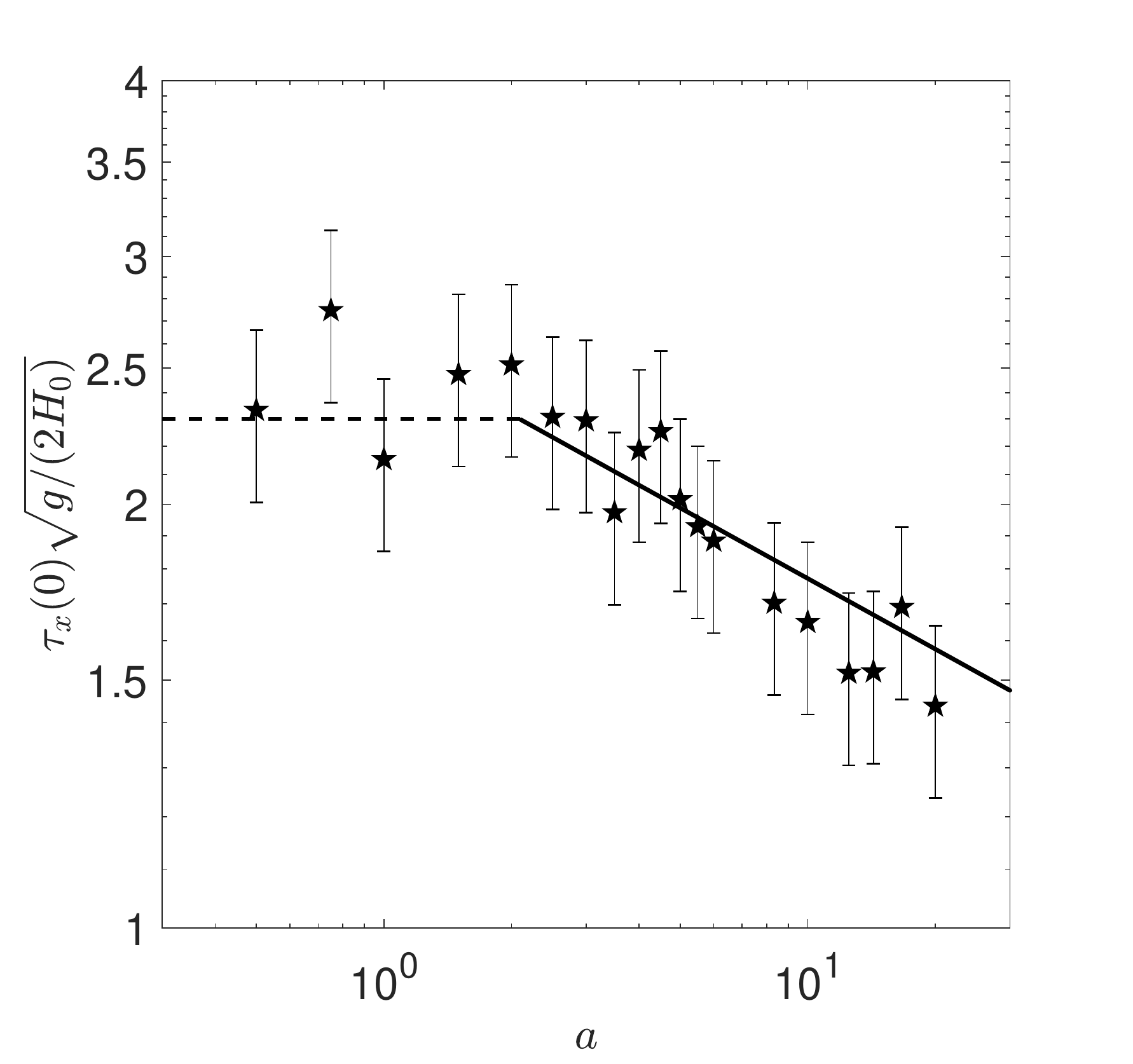}
	\caption{Evolution of the relative characteristic time of spreading at the base of the column $\tau_x(0)\sqrt{g/(2H_0)}$ as a function of the aspect ratio $a$. ($\bigstar$, with error bars) experimental data; (\dashedblack): $\tau_x(0)=2.3\sqrt{2H_0/g}$; (\fullblack): $\tau_x(0)=2.6a^{-1/6}\sqrt{2H_0/g}$.}
	\label{fig:fig8}
\end{figure}

\noindent fits well the data, with a prefactor again in agreement with the value $1/\sqrt{\overline{\beta}} \simeq 1.82$, which means that the characteristic time of spreading $\tau_x(z)$ at an altitude $z$ is also proportional to the free-fall like time over the corresponding typical lateral extension $\Delta L(z,\tau_x)$. 
Moreover, the inset in figure \ref{fig:fig7}(b) compares the final runout distance $\Delta L_f$ to the spreading length $\Delta L(0,\tau_x)$ at the base of the column at time $\tau_x(0)$. It should be pointed out that here again $\Delta L(0,\tau_x)$ and $\tau_x(0)$ are evaluated at $z=d$ to reduce measurement uncertainties. A linear relation of slope 1.43 can be inferred between the two characteristic lengths. Using this observation, we obtain from equation \eqref{eq_tau_x_z} the following scaling for the time of collapse at the base of the column:

\begin{equation}
	\tau_x(0) \propto \sqrt{2\Delta L_f/g}.
  \label{eq_tau_z_x}
\end{equation}

\noindent In addition, considering the expressions provided by equation \eqref{lajeunesse_runout} for the runout length $\Delta L_f$ we obtain

\begin{equation}
	\tau_x(0) \propto \left\{ \begin{array}{ll}
	\displaystyle \sqrt{2H_0/g}\\[8pt]
	\displaystyle a^{-1/6} \sqrt{2H_0/g}
	\end{array}\right.
    \begin{array}{ll}
	\displaystyle \ \mathrm{for\ } a \lesssim 3,\\[8pt]
	\displaystyle \ \mathrm{for\ } a \gtrsim 3.
	\end{array}
  \label{tau_x_0}
\end{equation}

\noindent Hence, for low aspect ratios ($a \lesssim 3$), the characteristic time of spreading should be proportional to the free-fall time $\sqrt{2H_0/g}$ over the initial height $H_0$, in agreement with the findings of Lajeunesse \textit{et al.} \cite{2005_lajeunesse}. However, there should also be a subtle influence of $a$ on the spreading time for high aspect ratios ($a \gtrsim 3$), which could explain the behavior reported by Lacaze \textit{et al.} \cite{2008_lacaze} [see figure (8) therein].\\
The characteristic spreading time at the bottom of the column, $\tau_x(0)$, normalized by $\sqrt{2H_0/g}$, is shown as a function of the aspect ratio in a log-log representation in figure \ref{fig:fig8}. A quite good agreement is observed with the scalings of equation \eqref{tau_x_0}, despite some scattering. In particular, a plateau value of about 2.3 is obtained, for $a$ lower than a critical value comprised between 2 and 3, while a slight decrease, beyond the error range, occurs at larger values of $a$, which is compatible with a power law of exponent $-1/6$. These observations highlight the influence of the aspect ratio of the column on the characteristic time of the granular spreading.

\subsection{Quasi-static regime}
\label{SubsecIVB}


For a low release velocity of the granular column, \textit{i.e.}, corresponding to the experiments of series B, the inertia of the grains is negligible, so that the granular slide exhibits a quasi-static evolution. In figures \ref{fig:fig4}(a) and \ref{fig:fig4}(b), the relative final height $H_f/L_0$ and runout distance $\Delta L_f/L_0$ are shown for these experiments (\textcolor{red}{$\blacksquare$}), with both parameters presenting a strict growth with the initial aspect ratio of the column. In figure \ref{fig:fig4}(a), the relative final height is found to either coincide (when $a \lesssim 0.9$) or be higher (for $a \gtrsim 0.9$) than the values obtained at large release velocity (series A). In figure \ref{fig:fig4}(b), the runout distance obtained for experiments at low release velocity is systematically lower than for experiments at large release velocity. As already mentioned in section III, a triangular shape with a straight slope is obtained experimentally for high enough aspect ratios ($a \gtrsim 0.9$), as observed in figure \ref{fig:fig2}(l). Considering the conservation of mass, and that the final geometry is described by the angle of repose $\theta_r$ of the granular material leads to

\begin{subequations}
\label{eq:lrs_ha} 
\begin{align}
\displaystyle H_0L_0 & =\frac{H_fL_f}{2}, \label{eq:lrs_haa}
\\
\displaystyle \frac{H_f}{L_f} & =\tan\theta_r. \label{eq:lrs_hab}
\end{align}
\end{subequations}

\noindent Solving equation (\ref{eq:lrs_ha}) in terms of $H_f$ and $L_f$ leads to the following expressions

\begin{subequations}
\label{eq:lrs_ha_scalings} 
\begin{align}
\displaystyle \frac{H_f}{L_0} & =\sqrt{2a\tan\theta_r}, \label{eq:lrs_ha_scalingsa}
\\
\displaystyle \frac{\Delta L_f}{L_0} & =\sqrt{\frac{2a}{\tan\theta_r}}-1. \label{eq:lrs_ha_scalingsb}
\end{align}
\end{subequations}


\noindent In contrast, for low enough aspect ratios, namely when $a \lesssim 0.9$, a trapezoidal shape is obtained, as part of the initial column remains at rest during the collapse [see figure \ref{fig:fig3}(j)]. Hence, using once again mass conservation and the fact that in this case  $H_f = H_0$, we obtain the following relations

\begin{subequations}
\label{eq:lrs_la_scalings} 
\begin{align}
\displaystyle \frac{H_f}{L_0} & =a, \label{eq:lrs_la_scalingsa}
\\
\displaystyle \frac{\Delta L_f}{L_0} & =\frac{a}{2\tan\theta_r}. \label{eq:lrs_la_scalingsb}
\end{align}
\end{subequations}


\noindent The critical aspect ratio $a_c$, which separates the triangular shape regime from the trapezoidal one, is straightforwardly derived from the equality of equations \eqref{eq:lrs_ha_scalingsa} and \eqref{eq:lrs_la_scalingsa}:

\begin{equation}
\label{eq:acr}
a_c=2\tan\theta_r.
\end{equation}

\noindent The predictions given by equations (\ref{eq:lrs_ha_scalingsa}), (\ref{eq:lrs_ha_scalingsb}), (\ref{eq:lrs_la_scalingsa}), and (\ref{eq:lrs_la_scalingsb}) are reported in figure \ref{fig:fig4}. An excellent agreement is obtained between these scalings and experiments made at low release velocity, for $\theta_r = 23.5^\circ$. This value of $\theta_r$ is consistent with the measured angle of repose of the spherical glass beads used in the present study. It also gives the following estimate for the critical aspect ratio: $a_c \simeq 0.87$, which corresponds quantitatively to the observed transition between trapezoidal and triangular shapes around 0.9. \\
It should be mentioned that these scalings for the quasi-static regime are identical to those obtained by Rondon \textit{et al.} \cite{2011_rondon}, or by Bougouin and Lacaze \cite{2018_bougouin}, for the collapse of an initially dense granular column in a viscous fluid. This is not surprising, as inertia is also negligible in the viscous regime, so that here again the collapse is quasi-static and the final morphology is governed by the angle of repose of the material. It is also in very good quantitative agreement with the results of Mériaux for the geometry of the deposits \cite{2006_meriaux}. Indeed, Mériaux studied the quasi-static collapse of a granular column when the retaining wall is slowly removed horizontally. In the case of glass beads with a diameter between $600\ \rm{\mu m}$ and $850\ \rm{\mu m}$, the relative final height was found to be equal to $a$ when $a \lesssim 2$ and to $a^{0.45}$ when $a \gtrsim 2$, while the rescaled runout distance $\Delta L_f/L_0$ was equal to $a$ when $a \lesssim 2$ and to $1.3~a^{0.7}$ when $a \gtrsim 2$. These expressions give results of the same level of accuracy as equations (\ref{eq:lrs_ha_scalingsa}), (\ref{eq:lrs_ha_scalingsb}), (\ref{eq:lrs_la_scalingsa}), and (\ref{eq:lrs_la_scalingsb}) for the data at low release velocity, in the range of explored aspect ratios.

\begin{figure}[t!]
	\centering
	\includegraphics[width=\columnwidth]{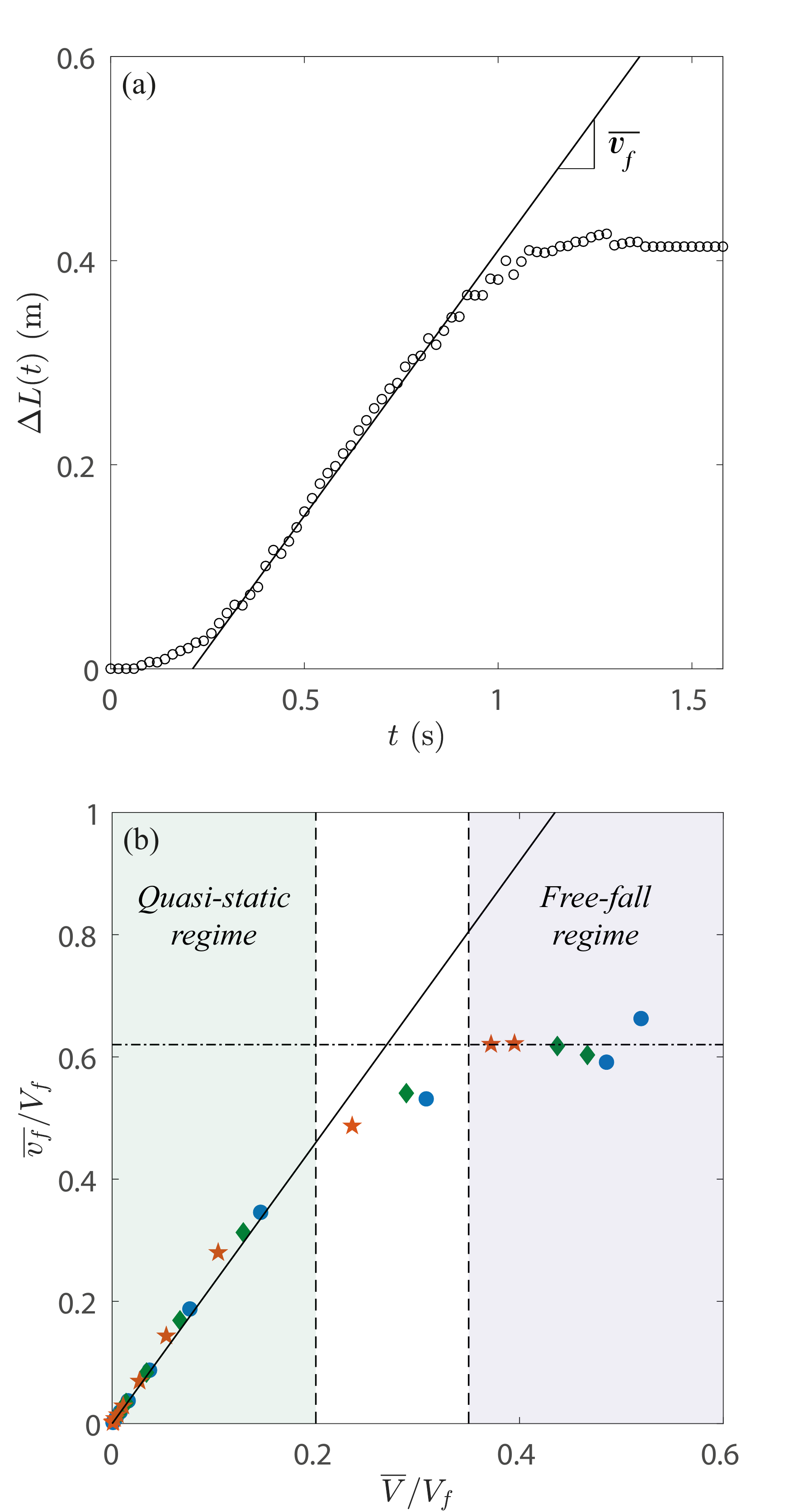}
	\caption{(a) Time evolution of the runout of a column with $H_0=37.5\ \rm{cm}$ , $L_0=7.5\ \rm{cm}$ ($a=5$), with a release velocity of $0.2\ \rm{m.s^{-1}}$; (\fullblack) estimation of the mean front velocity, with a slope $\overline{v_f}=0.52\ \rm{m.s^{-1}}$ in this case. (b) Evolution of the front velocity $\overline{v_f}$ with the mean release velocity $\overline{V}$, both non-dimensionalized by $V_f$. (\textcolor{darkblue}{$\bullet$}) $a=0.75$ ; (\textcolor{darkgreen}{$\blacklozenge$}) $a=2$; (\textcolor{darkorange}{$\bigstar$}) $a=5$; The horizontal dash-dotted line shows the plateau value $\overline{v_f}/V_f=0.62$; The continuous line corresponds to equation \eqref{eq:quasistatic_velocity_law} with $\theta_r = 23.5^\circ$; The vertical dashed lines highlight the values $\overline{V}/V_f \simeq 0.2$ marking the limit of the quasi-static regime, and	$\overline{V}/V_f \simeq 0.35$ above which no significant influence of the release velocity is observed.}
	\label{fig:fig9}
\end{figure}

\subsection{Influence of the release velocity}
\label{SubsecIVC}

The previous subsections characterized the two limiting cases of the free-fall and the quasi-static regimes, considering a high or low enough velocity of the sliding gate, respectively. A transition is expected between  these two asymptotic behaviors,  where the release velocity $V$ has still an impact on the collapse dynamics, and in particular on the velocity of the granular front $v_f$ at the base of the column.\\
To finely study the influence of the sliding gate, we compare the mean value of the release velocity $\overline{V}$, integrated over the time during which the gate is in contact with the grains, to the mean velocity $\overline{v_f}$ of the advancing granular front taken at the foot of the column, \textit{i.e.}, at $z=d$. The value of $\overline{v_f}$ is determined here as in Lajeunesse \textit{et al.} \citep{2005_lajeunesse}, \textit{i.e.}, by taking the tangent of the runout distance as a function of time within the region of nearly constant velocity, as illustrated in figure \ref{fig:fig9}(a) for an intermediate release velocity (nominal value $V=0.2\ \rm{m.s^{-1}}$, mean value $\overline{V}\simeq0.19\ \rm{m.s^{-1}}$). Even though the acceleration stage of the granular front at high release velocities was found to be quadratic in time in section  IV.A., this is not the case for the quasi-static regime. Indeed, for low release velocities the advancing front is roughly triangular, with a shape dictated by the angle of repose of the material. Hence, the definition of the mean front velocity $\overline{v_f}$ gives a more general indication of the collapse dynamics for the present discussion.\\
In figure \ref{fig:fig9}(b), we report for all data of series C-E  the evolution of $\overline{v_f}$ as a function of $\overline{V}$, both normalized by the typical advancing front velocity $V_f$ of the free-fall regime, 

\begin{equation}
	V_f=\sqrt{2\overline{\beta} g\Delta L_f} \simeq \left\{ \begin{array}{ll}
	\displaystyle 0.74~\sqrt{2gH_0}\\[8pt]
	\displaystyle 0.89~a^{-1/6} \sqrt{2gH_0}
	\end{array}\right.
    \begin{array}{ll}
	\displaystyle \ \mathrm{for\ } a \lesssim 3,\\[8pt]
	\displaystyle \ \mathrm{for\ } a \gtrsim 3,
	\end{array}
  \label{eq:Vff}
\end{equation}

\noindent where $\overline{\beta} \simeq 0.30$. The release velocity was varied over three decades in these experiments, for three representative initial aspect ratios of 0.75 (\textcolor{darkblue}{$\bullet$}), 2 (\textcolor{darkgreen}{$\blacklozenge$}) and 5 (\textcolor{darkorange}{$\bigstar$}), respectively.\\
All data collapse well onto a master curve, showing the same behavior at all considered aspect ratios. For $\overline{V}/V_f \lesssim 0.2$, the velocity of the advancing granular front $\overline{v_f}$ is governed by the release velocity, and a quasi-static evolution is observed for the granular slide. As discussed in section III, in this case, the moving interface exhibits at leading order a triangular shape dictated by the angle of repose of the material. Hence, during the spreading of the column, the runout distance, which can be approximated by $\Delta L(0,t) \simeq \overline{v_f}\,t$ as illustrated in figure \ref{fig:fig9}(a), is related to the distance $z_g(t) \simeq \overline{V}\,t$ from the bottom plane to the bottom of the gate through the relation

\begin{equation}
\label{eq:quasistatic_propagation_law}
\Delta L(0,t) \simeq \frac{z_g(t)}{\tan \theta_r},
\end{equation}

\noindent which, by taking the derivative of each terms, gives the following relation between $\overline{v_f}$ and $\overline{V}$

\begin{equation}
\label{eq:quasistatic_velocity_law}
\overline{v_f} \simeq \frac{\overline{V}}{\tan \theta_r}.
\end{equation}

\noindent This relation is represented by the continuous line in figure \ref{fig:fig9}(b), which fits well the experiments at low release velocity ($\overline{V}/V_f\lesssim 0.2$), with $\theta_r = 23.5^\circ$.
When $\overline{V}/V_f \gtrsim 0.35$, $\overline{v_f}/V_f$ saturates at a constant value of about $0.62 \pm 0.05$. Therefore, the value of $\overline{V}$ does not influence the granular collapse dynamics anymore, which corresponds to the free-fall regime. Combining this plateau value of 0.62 with equation \eqref{eq:quasistatic_velocity_law}, the transition between these two asymptotic regimes is expected around $\overline{V} \simeq 0.62\tan \theta_r V_f \simeq 0.3~V_f$. The saturation is clearly observed when $\overline{V}/V_f \gtrsim 0.35$, which gives a condition to prevent any significant effects of the release velocity on the collapse dynamics. Indeed, for the material used in the present study, namely spherical glass beads, a conservative criterion using equation \eqref{eq:Vff} would read

\begin{equation}
\label{eq:criterion_gb5mm}
\overline{V} \geq \gamma\sqrt{gH_0}.
\end{equation}

\noindent with $\gamma \simeq 0.4$. The coefficient $\gamma$ may slightly depend on the characteristics of the material considered, for instance the angle of repose of the granular medium, but is independent of the aspect ratio of the initial column. 

\section{Conclusion}
\label{SecV}

In the present paper, we report experimental results on the collapse of a dry granular column, where the velocity at which the grains are released is controlled. The aim was to get insights into the collapse dynamics and its influence on classical parameters such as the final height or the runout distance of the final deposit. The release velocity was varied over three decades, for several initial aspect ratios of the column. Different regimes for the collapse were identified depending on the release velocity~$V$.\\
For large values of $V$, classical power laws are recovered for $H_f$ and $\Delta L_f$ as a function of the initial aspect ratio $a$ \cite{2005_lajeunesse}. In this regime, the deposits exhibit a significant curvature. For high enough aspect ratios ($a \gtrsim 3$), the top of the granular column undergoes an overall free-fall like motion at a typical acceleration smaller than gravity, of about $0.5g$, followed by a deceleration to the final state. During the spreading of the collapsing column, the grains moving horizontally also experience a ``free-fall" like motion with a smaller acceleration, of about $0.3g$, followed by a deceleration stage leading to the final state of the deposits. By focusing on the bottom of the column, the duration of the granular collapse is found to either be constant at $a \lesssim 3$ or to slightly depend on the aspect ratio of the column when $a \gtrsim 3$, through a power law of exponent $-1/6$. This result explains the behavior reported in \cite{2008_lacaze}.\\
At low release velocities, a quasi-static evolution for the collapse is observed, where the motion is mainly controlled by the angle of repose of the material. In this case, the final shape of the deposits is triangular for high enough aspect ratios ($a \gtrsim 0.9$), and trapezoidal otherwise ($a \lesssim 0.9$), as part of the column remains strictly static. Based on these observations and using the conservation of mass, expressions for the final height $H_f$ and runout distance $\Delta L_f$ as a function of $a$ are derived, and the critical aspect ratio $a_c$ separating the triangular and the trapezoidal shapes is found to depend only on the angle of repose of the material. These scalings are found to be identical to those obtained for the viscous regime of the collapse of immersed granular columns \cite{2018_bougouin}, and quantitatively match the empirical scalings of Mériaux \cite{2006_meriaux}. In that respect, they seem to be characteristic of a quasi-static granular collapse, occurring once inertia is negligible.\\
Between these two asymptotic regimes at high and low release velocity, a transition exists, where an increasing release velocity has a decreasing influence on the collapse dynamics. In the present study, no more effect of the mean release velocity $\overline{V}$ on the collapse dynamics is observed as soon as $\overline{V} \gtrsim 0.35~V_f$, where $V_f$ is the typical advancing front velocity of the free-fall regime. This relation gives a practical criterion that should be used to ensure that the release process has no influence on the collapse dynamics in future experimental works. For the present investigation, in which glass beads were used, this criterion is $\overline{V} \geq 0.4\sqrt{gH_0}$.\\ 
In this study, we focused on quasi-two-dimensional granular collapses. However, it would be interesting to compare these results to the axisymmetric case, where different scalings govern the final morphology of the deposit \cite{2004_lajeunesse}. Besides, such a configuration is more realistic for describing large geophysical flows such as landslides, which are a threat for human facilities both in mountainous and coastal areas, as landslides entering into water are known for their tsunamigenic potential~\cite{2020_huang,2020_cabrera,2021a_robbe-saule,2021b_robbe-saule,2021_sarlin}.

\begin{acknowledgments}
The authors are grateful to J.~Amarni, A.~Aubertin, L.~Auffray and R.~Pidoux for the elaboration of the experimental setup.\\\\
\textbf{Declaration of interests.} The authors report no conflict of interest.
\end{acknowledgments}

\bibliography{2021_PRE}

\end{document}